\documentclass[journal]{IEEEtranTIE}
\usepackage{graphicx}
\usepackage{cite}
\usepackage{picinpar}
\usepackage{amsmath}
\usepackage{url}
\usepackage{flushend}
\usepackage[latin1]{inputenc}
\usepackage{colortbl}
\usepackage{soul}
\usepackage{multirow}
\usepackage{pifont}
\usepackage{color}
\usepackage{alltt}
\usepackage[hidelinks]{hyperref}
\usepackage{enumerate}
\usepackage{siunitx}
\usepackage{breakurl}
\usepackage{epstopdf}
\usepackage{pbox}
\usepackage[caption=false,font=footnotesize]{subfig}
\usepackage{array}
\usepackage{xcolor}

\begin{document}

\twocolumn[
\begin{@twocolumnfalse}

\begin{center}
\footnotesize{
This work has been submitted to the IEEE for possible publication. 
Copyright may be transferred without notice, after which this version may no longer be accessible.
}
\end{center}
\vspace{1em}

\title{Variable Dead-Time Based Novel Soft-Start Method for Dual Active Bridge Converters}

\author{
    \vskip 1em
    
    Sachith Wijesooriya, \emph{Graduate Student Member, IEEE},
    Sandun S. Kuruppu, \emph{Senior Member, IEEE}
    
    \thanks{
        Sachith Wijesooriya and Sandun S. Kuruppu are with the Electrical and Computer Engineering Department, Western Michigan University, Kalamazoo, 49008, USA.
    }
}

\maketitle

\end{@twocolumnfalse}
]
	
\begin{abstract}
Effective startup control is critical for the safe and reliable operation of Dual Active Bridge (DAB) converters. Unlike traditional soft-start techniques that rely solely on phase-shift control or fixed dead-time settings, the proposed approach gradually reduces the dead time from a value close to one switching period to the hardware-defined minimum. This enables a smooth buildup of the secondary-side voltage while effectively minimizing voltage overshoot and suppressing inrush current during startup. As a result, the leakage inductor current rises in a controlled manner, ensuring safe and predictable startup behavior. Simulation results demonstrate that conventional startup methods lead to severe voltage overshoot and high inrush currents, whereas the proposed method achieves a gradual voltage rise with well-regulated current profiles. Experimental validation using a 15 kW hardware platform confirms the effectiveness and robustness of the approach under different operating conditions. The proposed technique is simple, hardware-friendly, easily implementable on standard microcontrollers, and applicable to nth - order DAB architecture, making it a versatile solution for enhancing the reliability and safety of DAB converters in practical applications.
\end{abstract}


\section*{NOMENCLATURE}

\footnotesize
\begin{tabular}{ll}
$P_{\mathrm{DAB}}$ & Power transferred through the converter \\
$V_{\mathrm{dc}}$ & DC-link voltage (output voltage) \\
$V_{\mathrm{bat}}$ & Source or battery-side voltage \\
$C_{\mathrm{out}}$ & Output capacitance \\
$n$ & High-frequency transformer turns ratio \\
$L_{e}$ & Effective leakage inductance \\
$D$ & Normalized phase-shift ratio \\
$E_{c}$ & Energy stored in the output capacitor \\
$i_{o}$ & Output load current \\
$f_{s}$ & Converter switching frequency \\
$t_{d\_\mathrm{start}}$ & Initial dead time \\
$t_{d\_\mathrm{final}}$ & Final dead time \\
$T_{\mathrm{sw}}$ & Single switching period \\
\end{tabular}
\normalsize

{}

\definecolor{limegreen}{rgb}{0.2, 0.8, 0.2}
\definecolor{forestgreen}{rgb}{0.13, 0.55, 0.13}
\definecolor{greenhtml}{rgb}{0.0, 0.5, 0.0}

\section{Introduction}

\IEEEPARstart{A}{dvancements } in foundational technologies, materials, and control strategies are driving modern power electronics toward higher performance, increased power density, and superior efficiency, specifically tailored for emerging energy conversion applications \cite{ref1}. Sectors such as data center management \cite{ref2}, DC microgrids \cite{ref3}, fast EV charging \cite{ref4,ref5},\cite{ref6}, and advanced power distribution \cite{ref7,ref8},\cite{ref9,ref10} demand increasingly reliable and cost-effective energy conversion. Consequently, the Dual Active Bridge (DAB) has become a vital architecture, offering bidirectional power flow, galvanic isolation, and a compact magnetic profile enabled by high-frequency switching \cite{ref11}.

Additional benefits of a DAB include. 
\begin{itemize}
    \item Ability to operate efficiently over a wide voltage range while maintaining reduced semiconductor stress and low filtering requirements \cite{ref12,ref13}.
    \item Reduced switching losses enabled by zero-voltage switching (ZVS).
    \item Inherent voltage step-up/down capability allowing flexible voltage regulation \cite{ref14}.
\end{itemize}

While these advantages make the DAB a compelling solution, solving some of the known operational constraints remains critical. Addressing the challenges listed below will further improve its appeal and performance in diverse energy conversion sectors. Key challenges include,

\begin{itemize}
    \item Circulating currents under light-load conditions \cite{ref15}, leading to reduced efficiency.
    \item Limited ZVS operating range over wide voltage and load conditions, increasing switching losses.
    \item Increased conduction losses and electrical/thermal stress on the transformer and semiconductor devices due to high RMS currents and sensitivity to parameter variations \cite{ref16}.
    \item Limited transient handling capability, resulting in current and voltage spikes during dynamic events.
    \item Challenges in achieving proper soft start, where uncontrolled inrush current and voltage overshoot may occur.
\end{itemize}

Extensive research has addressed these limitations through the implementation of advanced modulation and control techniques designed to extend the operating range and maximize efficiency. Conventional methods such as single-phase shift (SPS), double phase shift (DPS) \cite{ref17}, extended-phase shift (EPS), and triple-phase shift (TPS) \cite{ref18} have been proposed to improve efficiency and expand soft-switching regions. More advanced approaches, including multi-phase shift modulation, asymmetrical duty-cycle control, variable-frequency operation, and hybrid modulation strategies \cite{ref19}, have further contributed to reducing circulating current and improving overall converter performance.

Among the remaining challenges, the implementation of an effective soft-start strategy remains a critical and actively researched topic for DAB converters \cite{ref20}, \cite{ref21}, \cite{ref22}. During startup, uncontrolled charging of the output capacitor and transformer magnetizing effects can lead to severe inrush current and voltage overshoot, potentially stressing semiconductor devices and passive components. Various soft-start strategies for DAB converters have been investigated in the literature, though each presents distinct limitations. For instance, the dual-phase-shift soft-start approach proposed in \cite{ref23} achieves a controlled startup sequence; however, its implementation complexity and the risk of significant current spikes on the primary side remain primary concerns. The technique described in \cite{ref24} manipulates gate signals exclusively during the initial switching period, which inherently limits its ability to govern the entire startup trajectory. While this approach provides immediate control, the charging dynamics of the output capacitor generally dictate the long-term voltage and current response; consequently, a single switching cycle is often insufficient to mitigate sustained inrush currents or prevent voltage overshoot. A more structured approach in \cite{ref25} employs a three-step soft-start strategy with inner phase-shift variation, experimentally illustrating the voltage and current evolution during startup. However, even in this case, the secondary bridge can experience an initial voltage buildup that induces in-rush current due to capacitor charging. Additional methods \cite{ref26} exploit secondary-side body diodes to gradually build output voltage before enabling phase-shift control, providing smoother startup. Similarly, a US patent \cite{ref27} proposes a dead time-based soft-start technique using a fixed dead time during startup followed by a minimum dead time in steady state, with duty-cycle variation from $15$ $\%$ to $60$ $\%$. 
While these approaches provide localized mitigation for specific startup issues, they generally lack the flexibility required to adapt to dynamic operating conditions. This limitation underscores the necessity for a more robust, scalable, and easily implementable control framework that ensures stability across a wide range of load and voltage variations.

To address these limitations, this paper proposes a variable dead-time based soft-start methodology for Dual Active Bridge converters. By dynamically modulating the effective duty ratio via dead-time adjustment during the power-up sequence, the proposed strategy effectively suppresses inrush current and eliminates voltage overshoot. The primary advantage of this method lies in its simplicity, providing high-performance transient management within the existing control framework. The remainder of this paper is organized as follows: Section II describes the operating principle of the DAB converter and the startup challenges. Section III presents the proposed soft-start strategy and its theoretical analysis. Section IV discusses simulation and experimental results validating the proposed method. Finally, Section V concludes the paper with key observations and future research directions.

\section{Dual active bridge overview and the startup challenges}
The Dual Active Bridge DC$-$DC converter enables bidirectional power transfer between two full bridges integrated with a high-frequency transformer, as illustrated in Fig. 1. The topology consists of two full-bridge circuits that generate duty-modulated square-wave voltages, $V_p$ and $V_s$, which are applied across a high-frequency transformer in series with an effective inductance $L_e$, \cite{ref28}. For analytical simplicity, all H-bridge switches are assumed to be ideal, and the magnetizing inductance of the transformer is neglected. The inductance $L_e$ represents the combined effect of the transformer leakage inductance and any external series inductance added to the circuit. Power transfer between the primary and secondary bridges is achieved by introducing a phase shift between $V_p$ and $V_s$.

\begin{figure}[htbp]\centering
	\includegraphics[width=8.5cm]{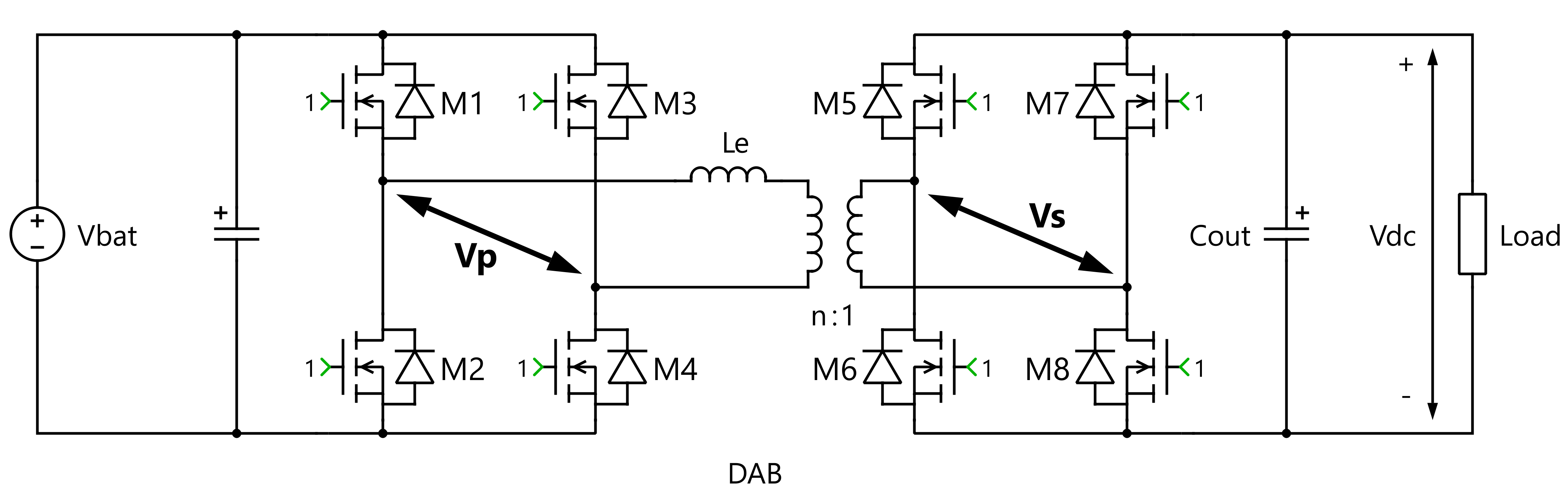}
	\caption{General overview of the DAB converter.}\label{FIG_1}
\end{figure}

In this study, the single-phase shift modulation scheme is employed to control the DAB. Under SPS control, both the primary and secondary bridges operate with a $50$ $\%$ fixed duty ratio, while the power flow and output current are regulated by adjusting the phase shift between the primary and secondary PWM signals. At startup, the abrupt application of a fixed $50$ $\%$ duty ratio presents significant operational challenges, primarily stemming from the transient DC offset in the transformer flux. Because the transformer begins at a zero-flux state, a full-width pulse can drive the magnetic core into saturation, leading to a sharp decline in inductive reactance and subsequent primary-side current spikes. Furthermore, since the output capacitor is initially discharged, this maximum energy transfer results in excessive inrush currents that can exceed the peak ratings of the power semiconductors. These high-current transients, coupled with the output filter's response, often trigger significant voltage overshooting and high-frequency ringing, potentially compromising component longevity and triggering over-current protection (OCP) circuits. Hence, a soft-start implementation, (regardless of the modulation scheme) is essential to mitigate these transient fluctuations, which are detrimental to system-level performance and long-term component reliability. Accordingly, a proper soft-start mechanism aims to address the following key cases.

\begin{itemize}
    \item Limiting Inrush Current during startup.
    \item Preventing Transformer DC Bias.
    \item Controlled Voltage Ramping and Pre-Charging.
\end{itemize}

Typically, many architectures, such as electric vehicle (EV) drivetrains, utilize a pre-charging stage to establish the DC-link voltage under no-load conditions. However, regardless of the load state, it is imperative that the startup sequence is regulated to prevent the aforementioned transients, specifically excessive inrush current and voltage overshoot, which can compromise system integrity. Figure 2 illustrates an example hard-startup behavior in the leakage inductor current and the output capacitor current. 

With these challenges, the proposed approach effectively suppresses inrush current and eliminates voltage overshoot across a wide input voltage range during the initial startup phase, thereby improving system reliability, operational safety as well as eliminating the need for a pre-charge circuit.

\begin{figure}[htbp]\centering
	\includegraphics[width=8.5cm]{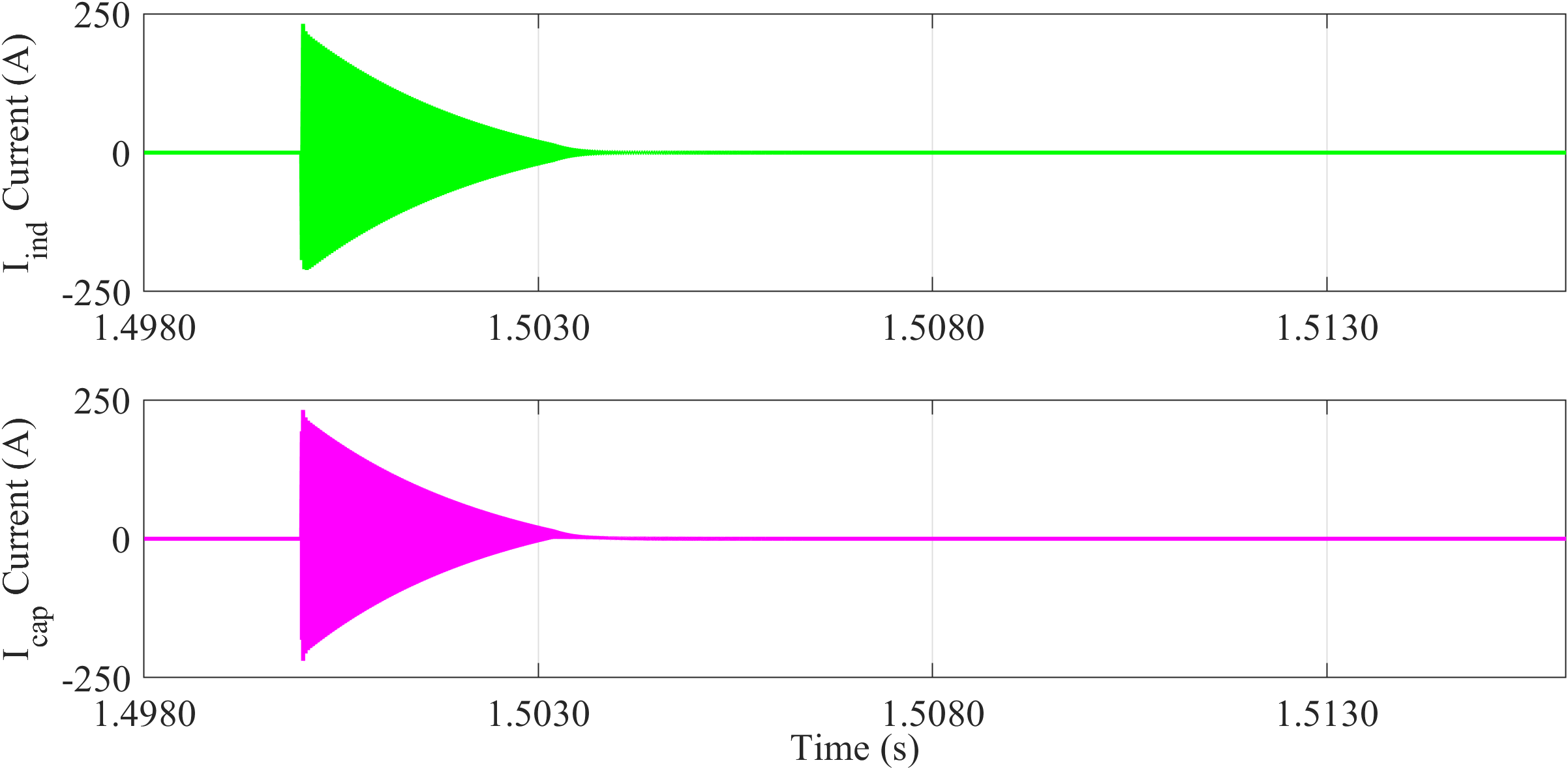}
	\caption{DAB capacitor and leakage inductor current behavior under normal operation at 650 V input voltage.}\label{FIG_2}
\end{figure}

\section{Proposed novel startup strategy}
The Dual Active Bridge converter enables bidirectional power flow control while providing galvanic isolation through a high-frequency transformer. The steady-state power transfer of the DAB can be expressed as (1),

\begin{align}
P_{\mathrm{DAB}} = 
\frac{V_{\mathrm{dc}} \, V_{\mathrm{bat}} \, D \left(1 - \left|D\right|\right)}
{2 n L_{e} f_{s}}
\end{align}

During startup, the DC-link capacitors on both the primary and secondary sides are usually uncharged or at relatively low voltages. If the converter is energized abruptly, the step-change in voltage across the transformer induces a significant inrush current to charge the output capacitor. Then, the energy stored in the output capacitor is given by (2),

\begin{align}
E_{C} = \frac{1}{2} C_{\mathrm{out}} V_{\mathrm{dc}}^{2}
\end{align}

The rate at which the capacitor absorbs energy can be expressed as (3),

\begin{align}
\frac{dE_{C}}{dt} = 
\frac{1}{2} \frac{d}{dt} 
\left( C_{\mathrm{out}} V_{\mathrm{dc}}^{2} \right)
\end{align}

By combining (1) and (3), the rate of change of the output voltage can be derived as (4),

\begin{align}
\frac{dV_{\mathrm{dc}}}{dt} =
\frac{V_{\mathrm{bat}} \, D \left(1 - \left|D\right|\right)}
{2 n L_{e} C_{\mathrm{out}} f_{s}}
-
\frac{i_{o}}{C_{\mathrm{out}}}
\end{align}

Multiplying both sides of (4) by $C_{out}$, the capacitor current can be expressed as (5),

\begin{align}
C_{\mathrm{out}} \frac{dV_{\mathrm{dc}}}{dt}
=
\frac{V_{\mathrm{bat}} \, D \left(1 - \left|D\right|\right)}
{2 n L_{e} f_{s}}
-
i_{o}
\end{align}

\begin{figure*}[!t]
\centering

\subfloat[]{
\includegraphics[width=0.48\textwidth]{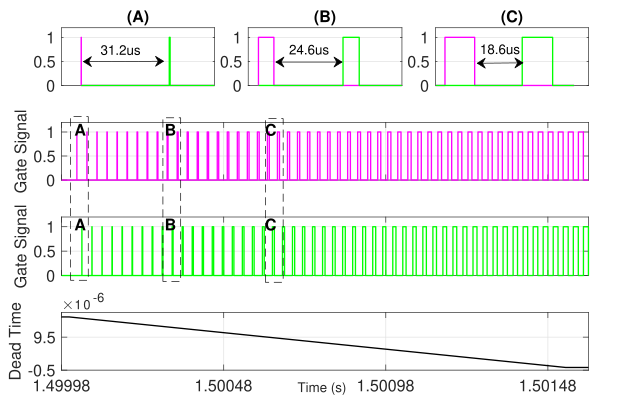}
\label{fig3a}
}
\hfill
\subfloat[]{
\includegraphics[width=0.48\textwidth]{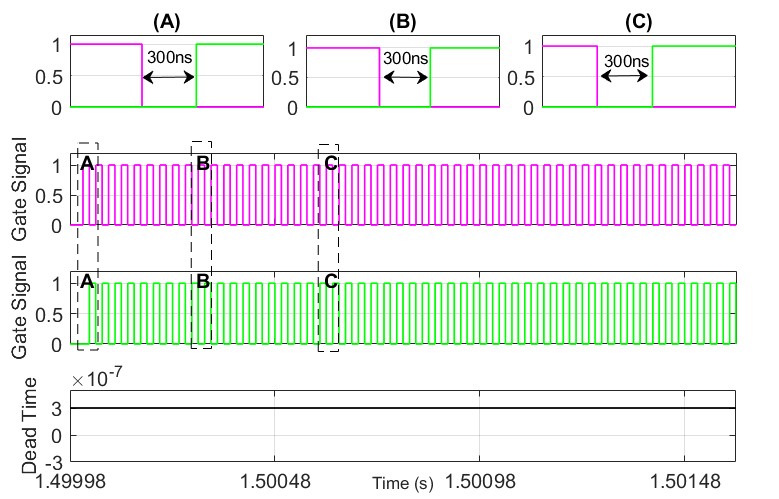}
\label{fig3b}
}

\caption{Gate signal variation of the DAB during startup: 
(a) proposed method with variable dead time, and 
(b) conventional hard startup with fixed dead time 
(Magenta trace: gate signals of M1, M4, M5, and M8; 
Green trace: gate signals of M2, M3, M6, and M7).}
\label{FIG_3}

\end{figure*}

This relationship confirms that the output voltage buildup strongly depends on the applied duty ratio and the phase-shift conditions. A rapid increase in phase shift results in a large $dV_{dc}/dt$, which directly leads to voltage overshoot and inrush current.
To achieve a proper soft start, the rate of change of the output voltage must be carefully controlled over a defined time interval. Thus, to achieve this, the proposed method controls the overall dead-time window dynamically, which effectively limits the duty applying during startup. In general, the proposed method operates by initializing the gate signals with a large dead-time value and progressively reducing it over a predefined time interval. This intentional dead-time variation effectively modifies the duty ratio seen by the power stage converter. 
The associated dead-time modulation profiles, highlighting the differences between the proposed methods and a hard start, are further detailed in Fig. 3(a) and Fig. 3(b) respectively. Both figures (a) and (b) in Fig.3 illustrate the variation of the gate signals applied to the top (pink) and bottom (green) MOSFETs of the primary side half bridge HB1 for each of the respective methods. Third subplot of each sub figure in Fig. 3 shows the variation of dead time during the start-up process for each method. Note that, per the principle of operation of a DAB the same dead time values are uniformly applied to all other legs of the DAB converter, both on the primary and the secondary.
Fig. 4 illustrates the simulated performance of the DAB converter at each stage during both the conventional hard-start and the proposed soft-start sequences, obtained using the PLECS environment. The results demonstrate a significant reduction in the peak transients of both the leakage inductor current and the output capacitor current. This is accompanied by a gradual and regulated buildup of the secondary-side voltage. By subduing these transient oscillations, the proposed dead-time control strategy enhances the DAB converter's reliability and extends the operational lifespan of critical power components. Furthermore, this approach facilitates a seamless startup, providing broader system-level benefits such as reduced EMI and improved stability for downstream loads.

\begin{figure*}[htbp]
\centering
\includegraphics[width=18.1cm]{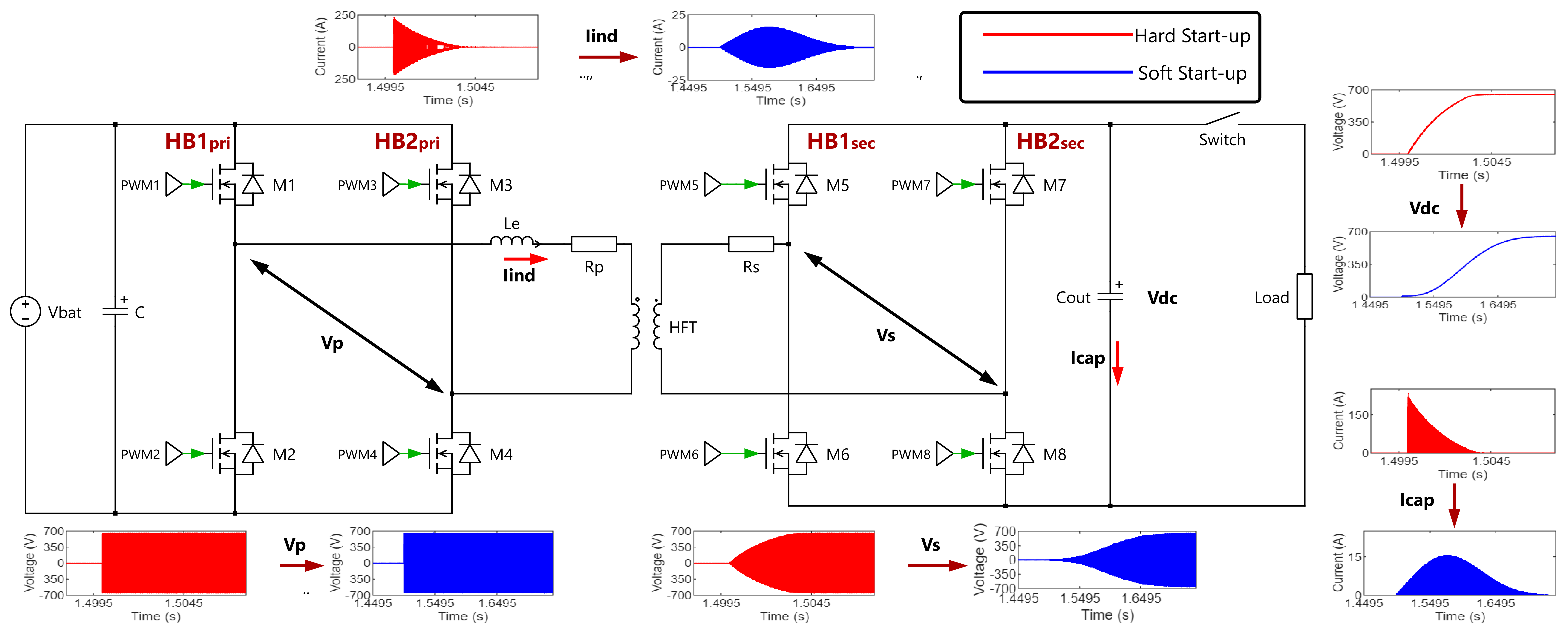}
\caption{DAB converter signal behavior under hard startup and the proposed soft startup methods.}
\label{FIG_4}
\end{figure*}

The following description further elaborates the proposed approach along with practical considerations. During normal operation, the DAB controller generates $50$$\%$ modulated PWM signals, which are applied to each switch. In conventional operation, a constant dead-time is maintained between the complementary PWM signals of each phase-leg to prevent shoot-through, a condition where both high-side and low-side switches conduct simultaneously, leading to catastrophic failure. The minimum dead-time requirement is determined by the specific switching characteristics of the power MOSFETs, such as turn-off delay and fall times, as well as the propagation delay mismatch inherent in the gate driver circuitry. Ensuring a sufficient safety margin is critical to maintain system integrity across varying temperature and load conditions. In the proposed methodology, the initial dead-time ($t_{d\_start}$) is configured to be a value approaching but not exceeding a single switching period ($T_{SW}$). This value is then linearly decreased to the nominal operational dead-time ($t_{d\_final}$) over a predefined time window, calibrated to the specific startup requirements of the application. The parameter $t_{d\_final}$ is defined by the standard dead-time requirements for steady-state system operation, ensuring a seamless transition from the soft-start phase to normal regulation. The duration of the startup window can be tailored to the application: a narrower window allows for a rapid, high-performance startup, whereas an extended window facilitates a smoother, more gradual transition. The primary advantage of employing a significant initial dead-time approaching the full switching period is that it keeps the secondary-side voltage near zero at the moment of energization. As the dead-time is linearly decreased, the energy transfer increases incrementally, enabling a regulated DC voltage buildup in the secondary side and mitigating potential output overshoots.

There are several key considerations for practical implementations of the proposed method, and  are discussed using the Piecewise Linear Electrical Circuit Simulation (PLECS) model shown in Fig. 5.

\begin{figure*}[htbp]
\centering
\includegraphics[width=18.1cm]{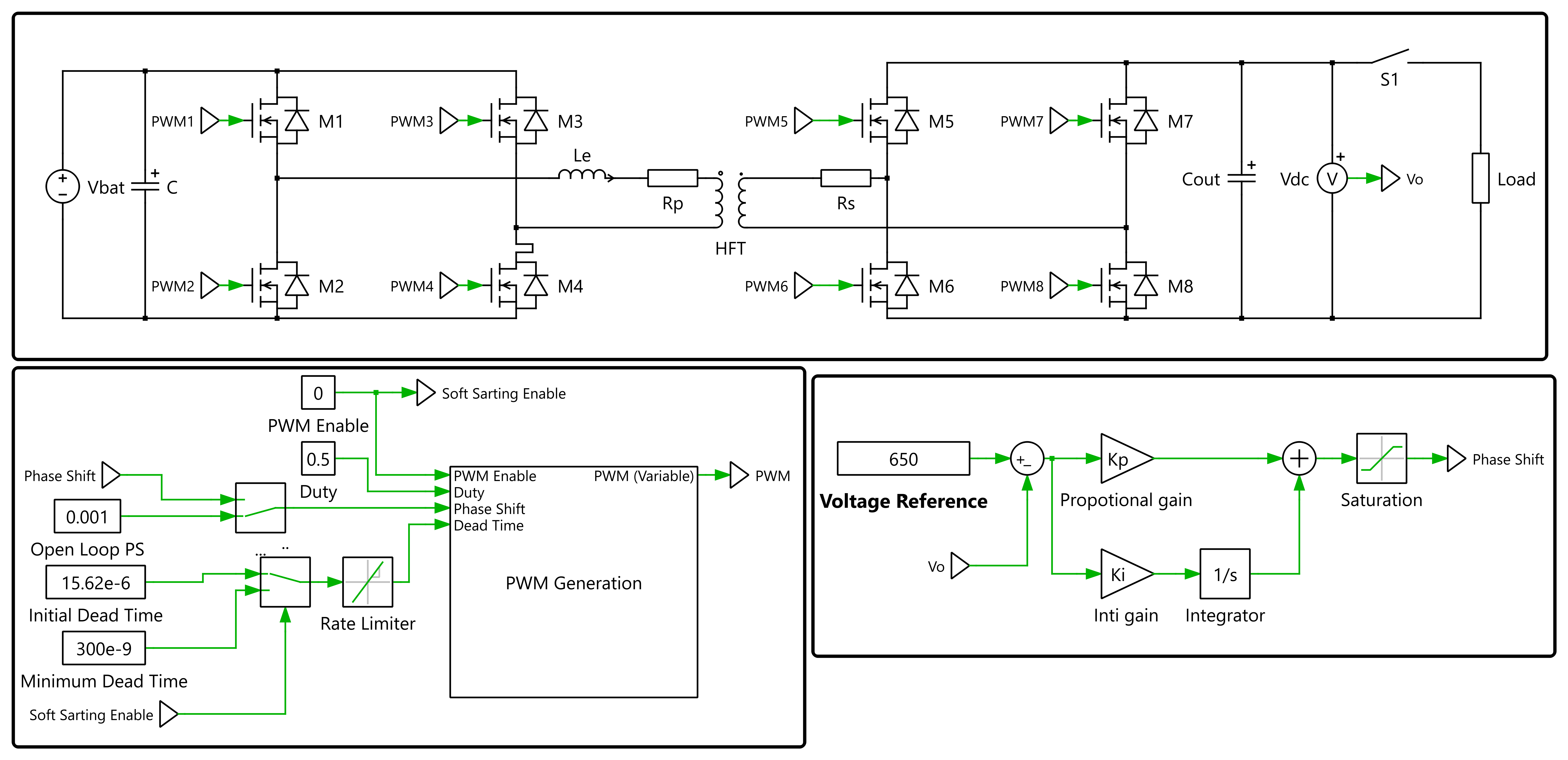}
\caption{Detailed controller architecture of the proposed soft starting method.}
\label{FIG_5}
\end{figure*}

The initial dead time $t_{d\_start}$ is selected based on the maximum achievable dead time of the system and is constrained to be less than one switching period. This selection also depends on the microcontroller clock frequency and the PWM generation configuration, which limits the minimal incremental time adjustment possible in realizing the maximum possible deadtime subjected to the specified constraint. 

In microcontroller-based implementations, dead time is typically introduced by adjusting the timing between two complementary PWM signals. Specifically, the insertion of dead time modifies both the falling-edge and rising-edge timings of the complementary PWM waveforms. Accordingly, the dead-time value programmed in the microcontroller is set to half of $t_{d\_start}$, which ensures that an equal amount of delay is applied to both the falling and rising edges of the PWM signals. The final dead-time value, $t_{d\_final}$, can be selected based on hardware constraints such as gate-driver characteristics, device switching behavior, and soft-switching requirements.
The rate of change of the dead-time during the soft-start interval is a configurable parameter, determined by the specific requirements of the DAB application and the broader system architecture. While sensitive loads may demand a prolonged, low-gradient ramp for maximum smoothness, other applications may prioritize a rapid transition to steady-state operation. The dynamic behavior of the DAB under various ramp-rate configurations is further detailed in the experimental section of this paper.
The following sections provide a comprehensive performance comparison using PLECS simulations, evaluating the proposed variable dead-time control against conventional hard-start and existing soft-start methodologies. Finally, the proposed method is validated through experimental results, confirming its effectiveness in real-world operating conditions.

\section{Simulation and experimental verification}
This section provides a comprehensive evaluation of the proposed variable dead-time control strategy through both PLECS-based simulations and experimental validation. To establish a rigorous benchmark, the performance of the proposed approach is compared against conventional hard-start sequences and several state-of-the-art soft-start methodologies, such as DPS-based duty-cycling and DC-bias suppression-based modulation
\subsection{Simulation verification}
Simulation analysis was conducted using the PLECS (Plexim) environment. A key advantage of this platform is the integrated TI C2000 Target Support Package, which includes an exclusive PWM signal generator block. This block accurately models the Texas Instruments (TI) microcontroller's hardware peripherals, enabling high-fidelity offline switching-level simulations that align closely with the embedded code's behavior. The Dual Active Bridge converter was designed to deliver a nominal power of 15 kW at a 650 V DC-link voltage. The system parameters utilized for both simulation and subsequent experimental validation are summarized in Table I.

\begin{table}[htbp]
\footnotesize
\renewcommand{\arraystretch}{1.3}
\caption{DAB Converter Parameters}
\label{table_dab}
\centering
\resizebox{\columnwidth}{!}{
\begin{tabular}{l l}
\hline\hline \\[-3mm]
\multicolumn{1}{c}{Parameter} & \multicolumn{1}{c}{Value} \\[1.6ex]
\hline
Rated Power & 15 kW \\
Battery Voltage & 650 V \\
DC-link Voltage & 650 V \\
High-Frequency Transformer Turns Ratio & 1:1 \\
Leakage Inductance & 22 $\mu$H \\
DC-Link Capacitance & 120 $\mu$F \\
DAB Switching Frequency & 32 kHz \\
\hline\hline
\end{tabular}
}
\normalsize
\end{table}

The transient response of the DAB converter under conventional hard-start and the proposed soft-start conditions is illustrated in Fig. 6(a) and (b), respectively. In the hard-start scenario, the leakage inductor current, transformer primary/secondary voltages, and the output voltage reach the nominal 650 V within a negligible timeframe, triggering a high-magnitude inrush current. The secondary-side capacitor current and voltage waveforms further confirm that, absent the soft-start procedure, the system is subjected to severe ${di}/{dt}$ and ${dv}/{dt}$ transients, which pose a risk of catastrophic component failure.

\begin{figure*}[!t]
\centering

\subfloat[]{
\includegraphics[width=0.48\textwidth]{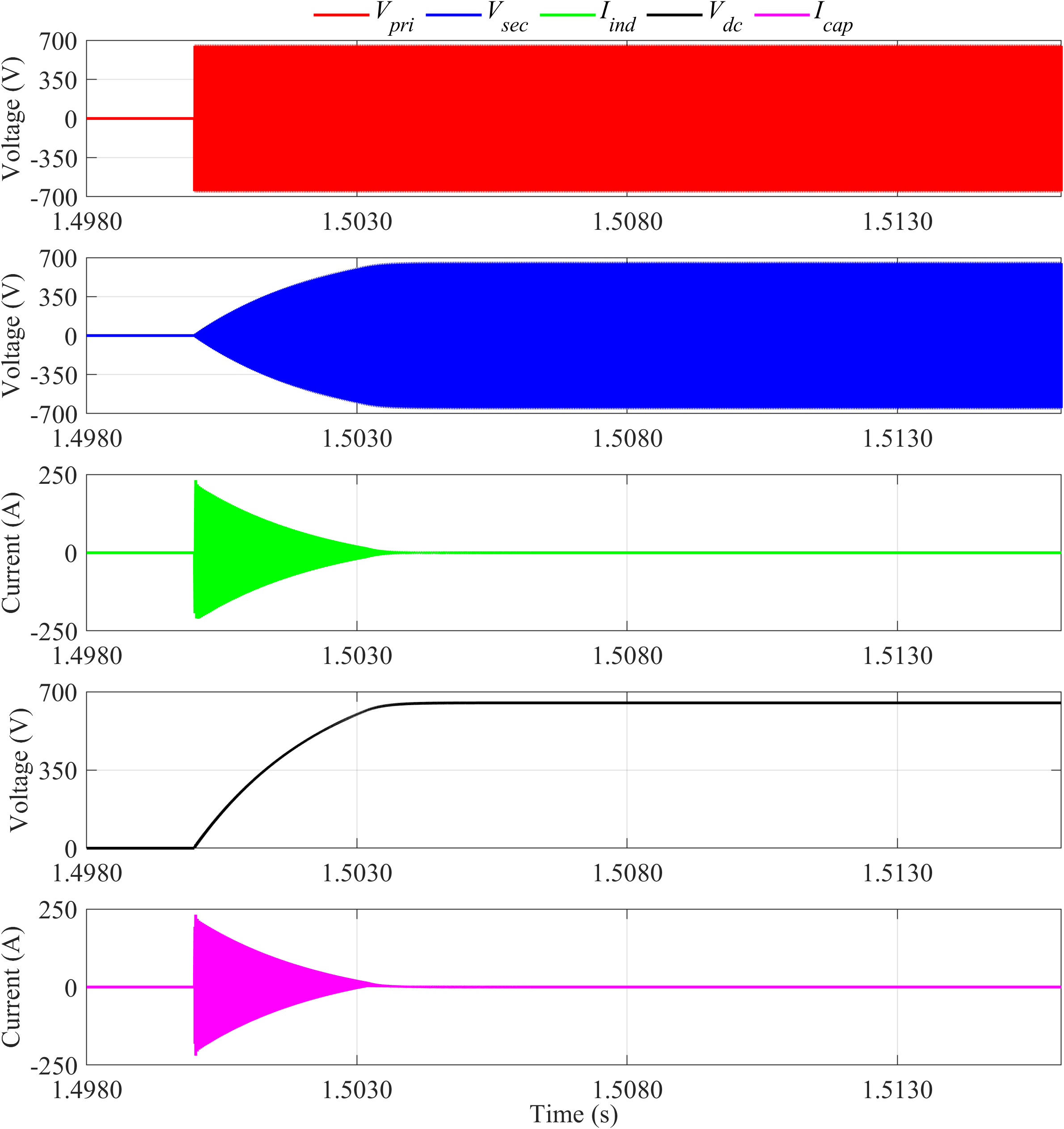}
\label{fig6a}
}
\hfill
\subfloat[]{
\includegraphics[width=0.48\textwidth]{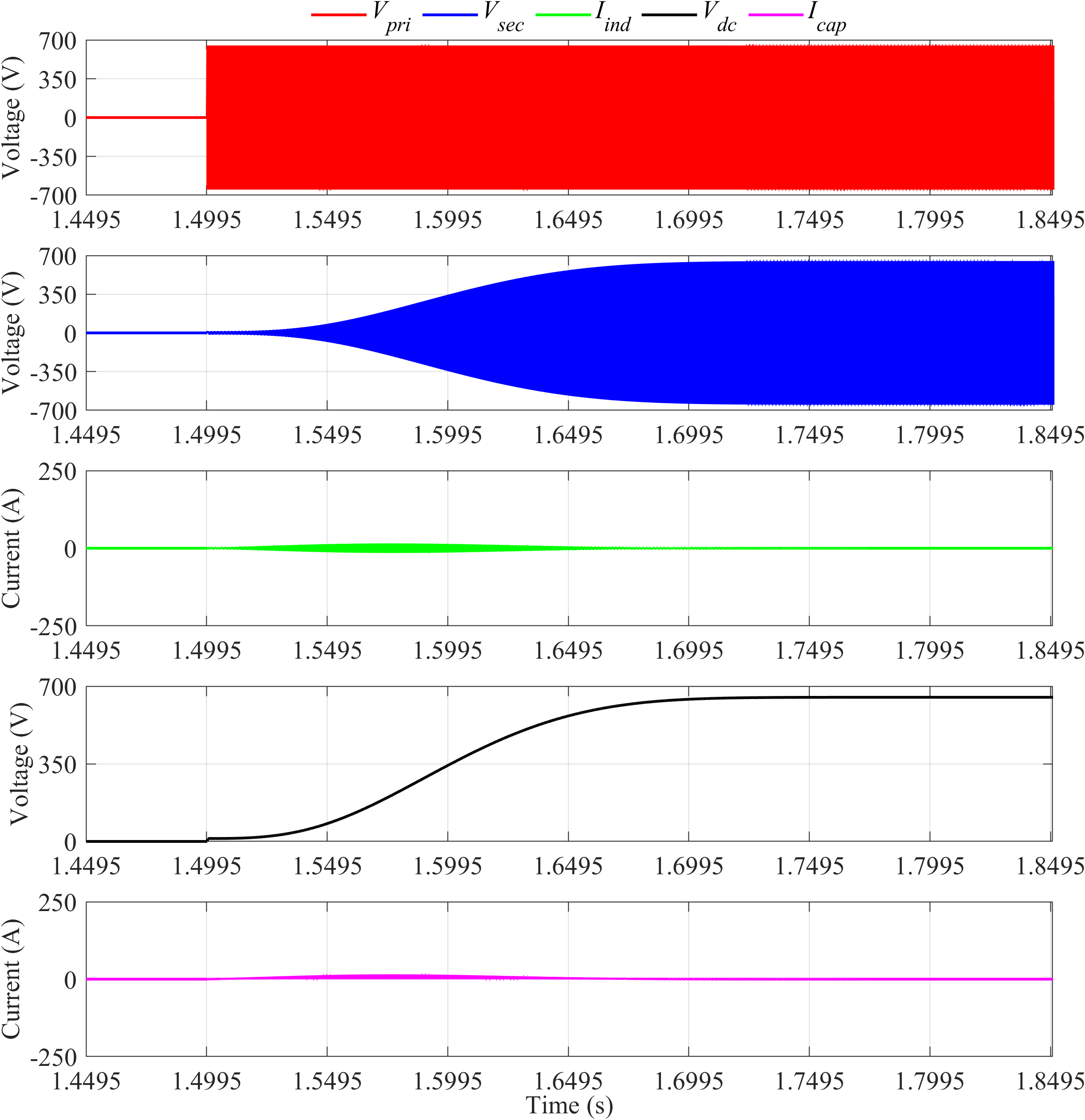}
\label{fig6b}
}

\caption{Startup behavior of the DAB converter showing 
(a) conventional hard startup and 
(b) the proposed variable dead-time-based soft-start method.}
\label{FIG_6}

\end{figure*}

Conversely, implementing the proposed variable dead-time method yields a marked improvement in the power stage profiles. As shown in Fig. 6(b), the secondary-side voltage increases monotonically in accordance with the dead-time variation, allowing the DC-link voltage to rise gradually without abrupt transitions. This regulated voltage escalation ensures a stable buildup of the leakage inductor current, effectively suppressing the high-amplitude spikes characteristic of unregulated startup sequences. 

The switching frequency of the DAB converter is 32 kHz, and the minimum dead time was set to 300 ns. The system PWM was enabled at $t$ = 1.5 s, and the initial dead time was set to 15.62 $\mu$s for both the rising and falling edges of the PWM carrier. This large initial dead-time ensures that the effective power transfer is minimal during the first stage of startup, allowing the converter to charge the secondary-side capacitor smoothly. The rate of change of the dead time is selected such that $t_{d\_start}$ transitions to $t_{d\_final}$ within 150 ms in this example.

\subsection{Comparison between the proposed method and existing methods}
This section provides a detailed comparison between representative soft-start methods reported in the literature and the proposed variable dead time based approach. To ensure a fair and consistent evaluation, the soft-start techniques presented in \cite{ref23}, \cite{ref24}, and \cite{ref27} were implemented and simulated in PLECS using an identical 15 kW DAB converter model and operating conditions. The resulting leakage inductor current and output DC-link voltage responses during startup are presented in Fig. 7 for each method, alongside the proposed method.

\begin{figure}[htbp]\centering
	\includegraphics[width=8.5cm]{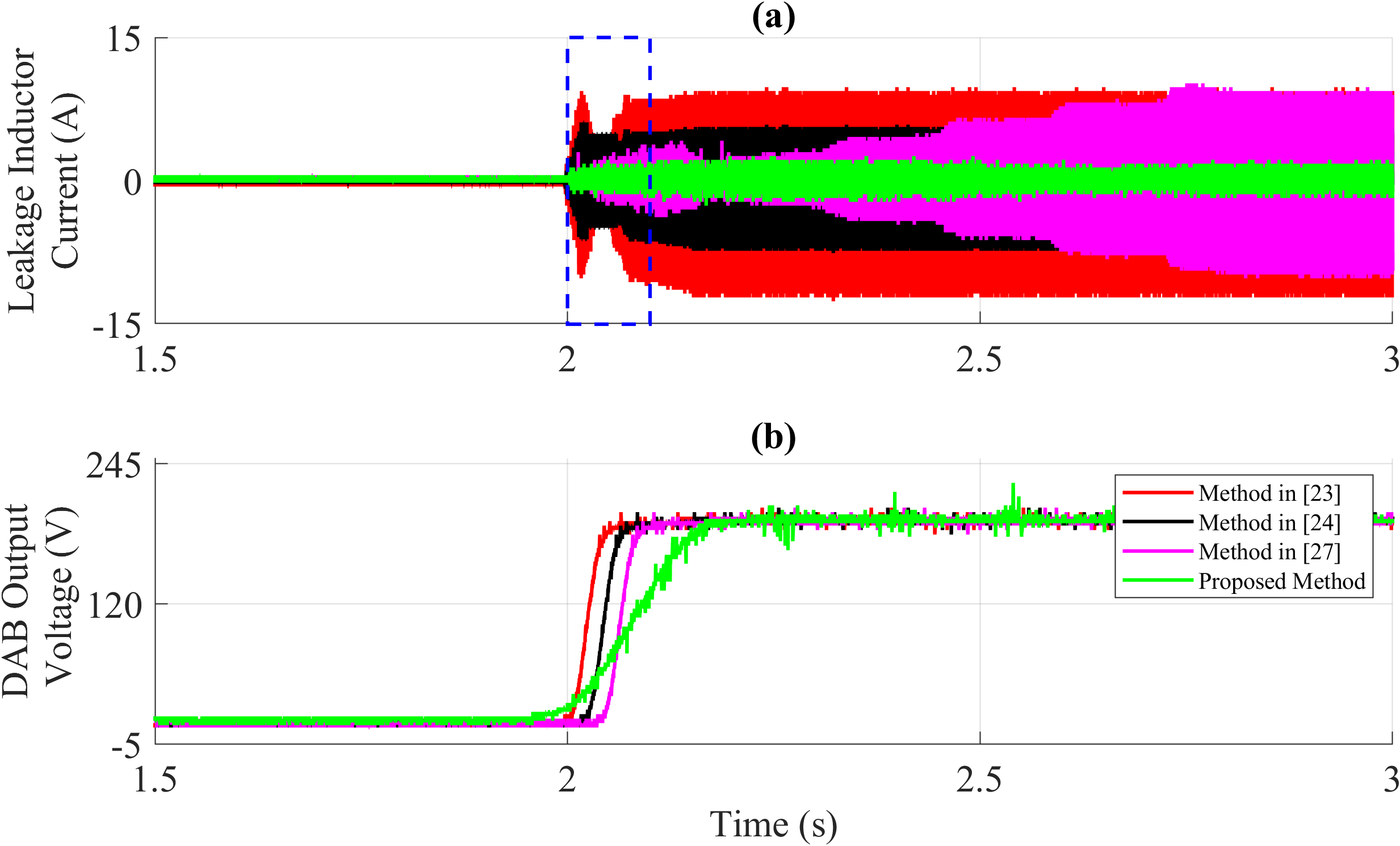}
	\caption{Comparison of soft-start methods with respect to the proposed novel dead-time control-based approach.}\label{FIG_7}
\end{figure}

The simulated waveforms closely replicate the behaviors reported in the original publications. Figure 8 shows a zoomed-in view of the leakage inductor current during the initial time interval indicated in Fig. 7 (blue dashed area). A comparative assessment of peak current stress, voltage overshoot tendency, control complexity, and implementation requirements are summarized in Table II, clearly highlighting the advantages of the proposed method in terms of controlled current and voltage buildup and reduced transient stress.

\begin{figure}[htbp]\centering
	\includegraphics[width=8.5cm]{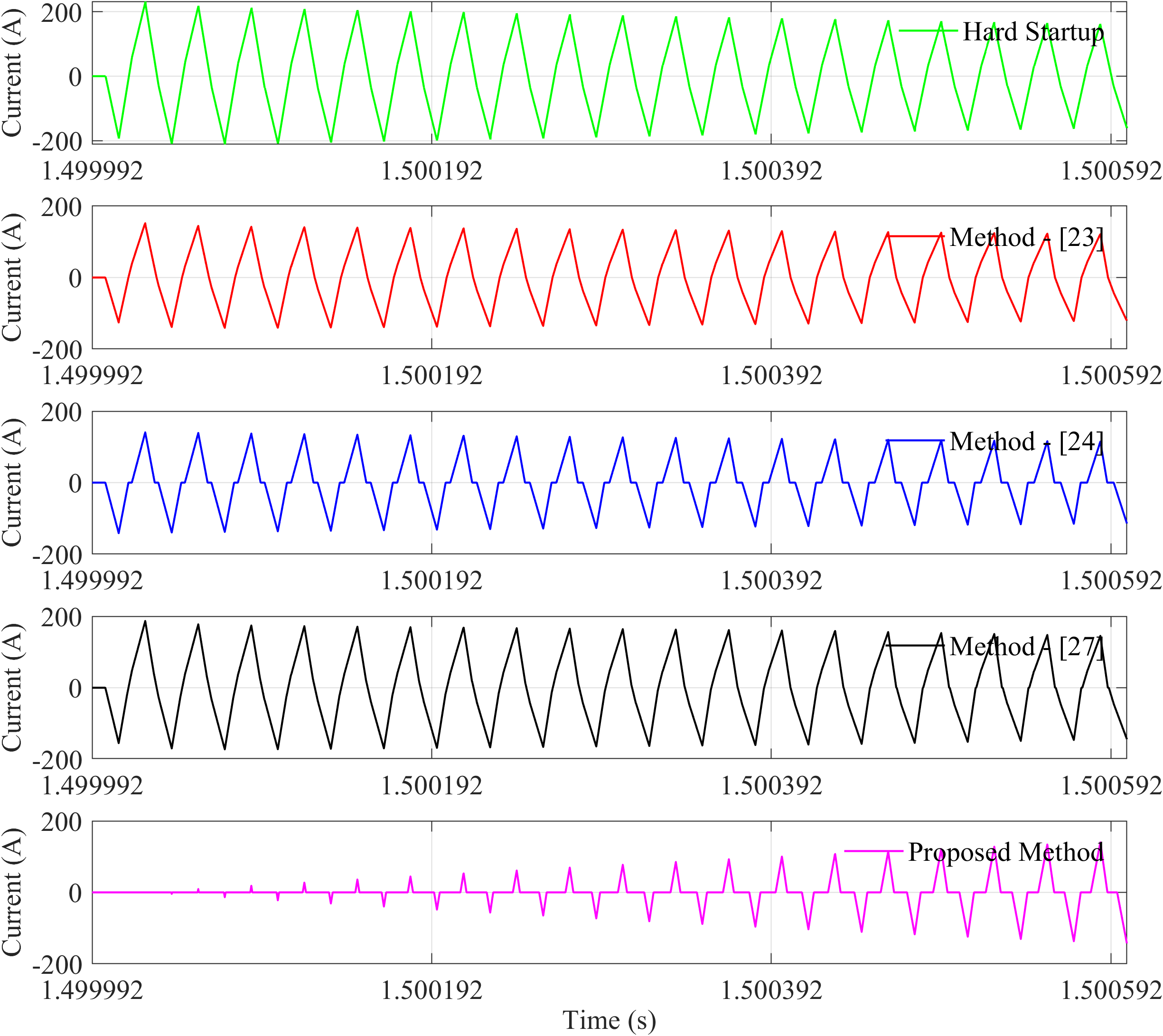}
	\caption{Comparison of leakage inductor current during startup for hard-start, existing soft-start methods, and the proposed approach.}\label{FIG_7}
\end{figure}

\begin{table*}[htbp]
\renewcommand{\arraystretch}{1.3}
\caption{Comparison of existing and proposed soft-start methods}
\label{table_softstart}
\centering
\begin{tabular}{p{1.3cm} p{3cm} p{3.1cm} p{2.5cm} p{1.5cm} p{4cm}}
\hline\hline 
\multicolumn{1}{c}{\shortstack{\textbf{Method}\\ \textbf{}}} &
\multicolumn{1}{c}{\shortstack{\textbf{Inductor Peak } \\\textbf{Current} }} &
\multicolumn{1}{c}{\shortstack{\textbf{Voltage }\\\textbf{Overshoot} }} &
\multicolumn{1}{c}{\shortstack{\textbf{Complexity}\\\textbf{} }} &
\multicolumn{1}{c}{\shortstack{\textbf{Startup Ramp} \\ \textbf{Controllability}}} &
\multicolumn{1}{c}{\shortstack{\textbf{Compatibility with} \\ \textbf{ 15 kW Model}}} \\
\hline

DPS-based soft-start \cite{ref23} 
& \textbf{Yes}; Lower than hard-start operation; however, noticeable transient current peaks occur during initial startup.
& \textbf{Yes}; comparable to the proposed method, slight overshoot during the transient interval. 
& \textbf{High}; due to dual phase-shift implementation and increased control complexity. 
& \vspace{0.3cm}\centering \textbf{No} 
& Original publication validated for low-voltage applications (25 V). Demonstrates high inductor current with the 15 kW simulation model.\\

DC-bias suppression using one switching cycle \cite{ref24} 
& \textbf{Yes}; Comparable to DPS-based methods; current variation reduces the initial transient magnitude.
& \textbf{Yes}; Initial voltage spike observed due to rapid output capacitor charging, as a single switching cycle is insufficient for full voltage buildup. 
& \textbf{Medium}; can be implemented within a standard digital control environment. 
& \vspace{0.3cm}\centering \textbf{No} 
& Original publication validated at medium voltage levels (180 V). Demonstrates relatively high inductor current with the 15 kW simulation model, but lower than the DPS based method.\\

Fixed large dead-time during soft-start interval \cite{ref27} 
& \textbf{No}; No sharp current spike during initial startup; however, a relatively high steady current flows due to the large duty ratio.
& \textbf{No}; ; No voltage transient during initial voltage buildup; however, a noticeable transient occurs when transitioning from large duty to nominal duty operation.
& \textbf{Low}; simple to implement in conventional control environments. 
& \vspace{0.3cm}\centering \textbf{No} 
& Original publication validated for low-voltage operation (60 V).
When applied to the 15 kW model, the transient becomes significantly pronounced.\\

Proposed variable dead-time based soft-start 
& \textbf{No}; Well-controlled and limited inductor peak current throughout startup, with no abrupt current spikes. 
& \textbf{No}; Voltage overshoot effectively eliminated; smooth and monotonic output voltage buildup. 
& \textbf{Low}; controller-friendly and easily implementable on standard microcontrollers. 
& \vspace{0.3cm}\centering \textbf{Yes} 
& Validated on the 15 kW model, demonstrates consistent and stable performance under high-voltage operation and practical system conditions. \\

\hline\hline
\end{tabular}
\end{table*}

The waveforms in Fig. 8 clearly demonstrate that, with the proposed soft-start method, the leakage inductor current increases gradually during the startup interval. Once the output capacitor is fully charged, the current smoothly decreases and settles close to zero, which is consistent with the trends observed in the previous simulation results.
\newpage
\subsection{Experimental results}

The proposed approach was experimentally validated with a SiC based 15kW DAB. The specifications of the DAB match that of the simulated system parameters given in Table I, and is shown in Fig. 9. The DAB includes CAB016M12FM3 SiC half-bridge modules driven by UCC21710-Q1 gate drivers. A TMDSCNCD280039C MCU control card was for PLECS based closed loop controller implementation and real-time control of the system.

\begin{figure}[htbp]\centering
	\includegraphics[width=8.5cm]{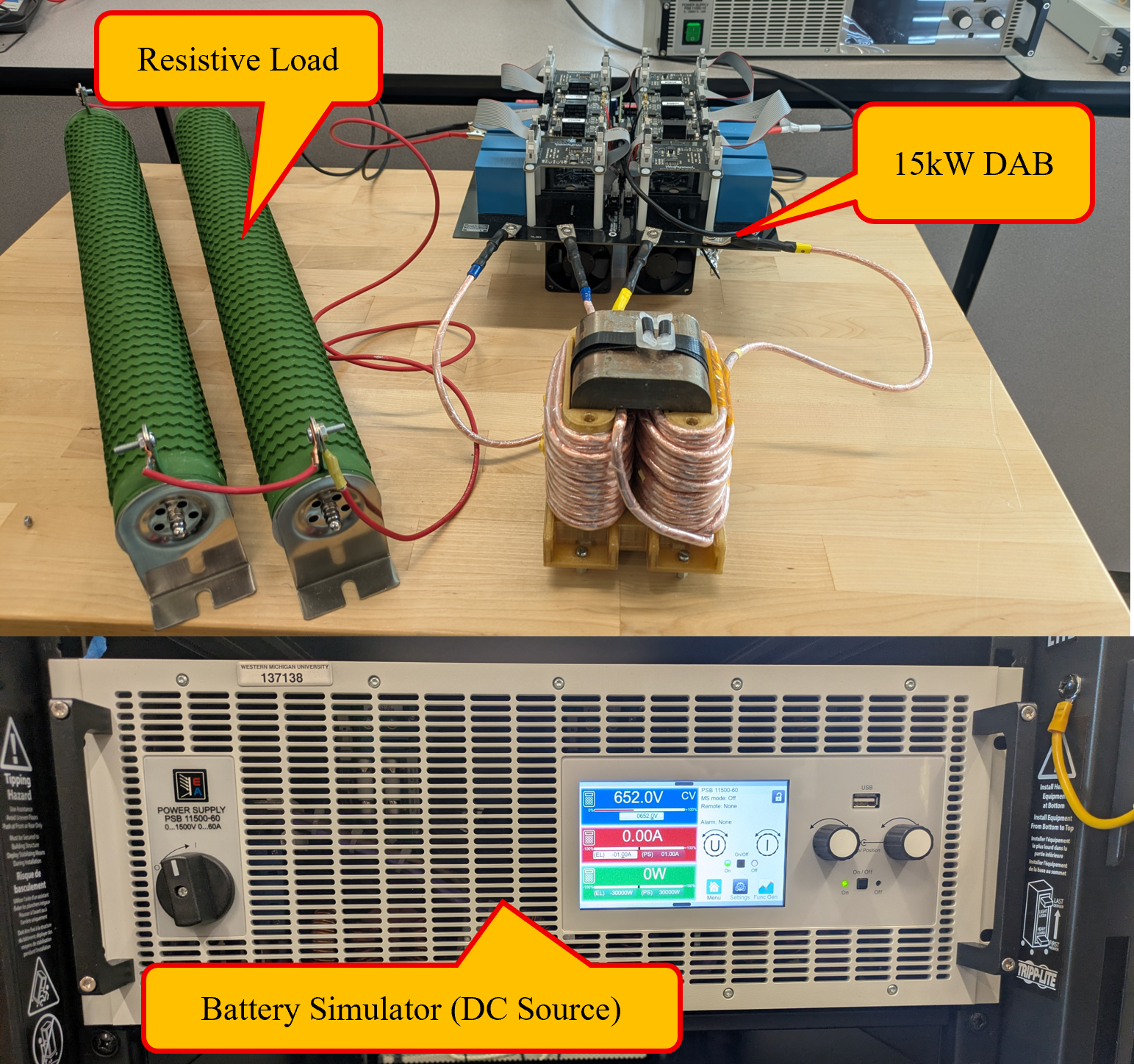}
	\caption{Experimental test setup is used for soft start method validation.}\label{FIG_9}
\end{figure}

To validate the robustness of the proposed soft-start strategy, experimental tests were conducted at three distinct DC-link voltage levels: 200 V, 400 V, and the rated 650 V. The corresponding experimental waveforms, captured via a high-resolution oscilloscope, are presented in Figs. 10, 11, and 12, respectively. The results demonstrate that the method maintains stable and consistent performance across this wide operating envelope. Specifically, the envelope of the leakage inductor current, the monotonic secondary-side voltage transition, and the regulated DC-link escalation all align with the predicted theoretical behavior throughout the soft-start interval.

\begin{figure}[htbp]\centering
	\includegraphics[width=8.5cm]{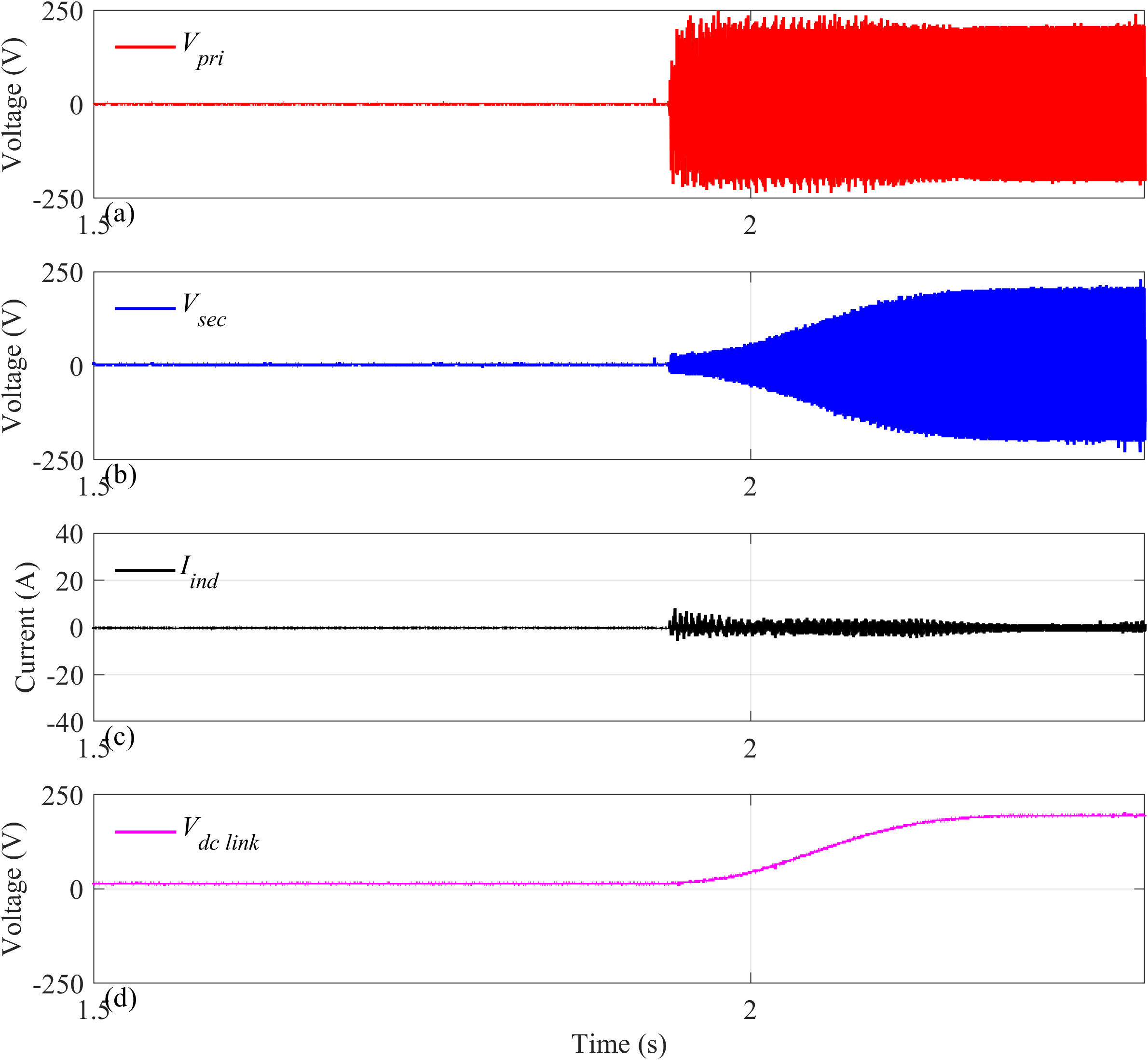}
	\caption{DAB behavior under proposed variable dead time-based startup process at 200V}\label{FIG_10}
\end{figure}

\begin{figure}[htbp]\centering
	\includegraphics[width=8.5cm]{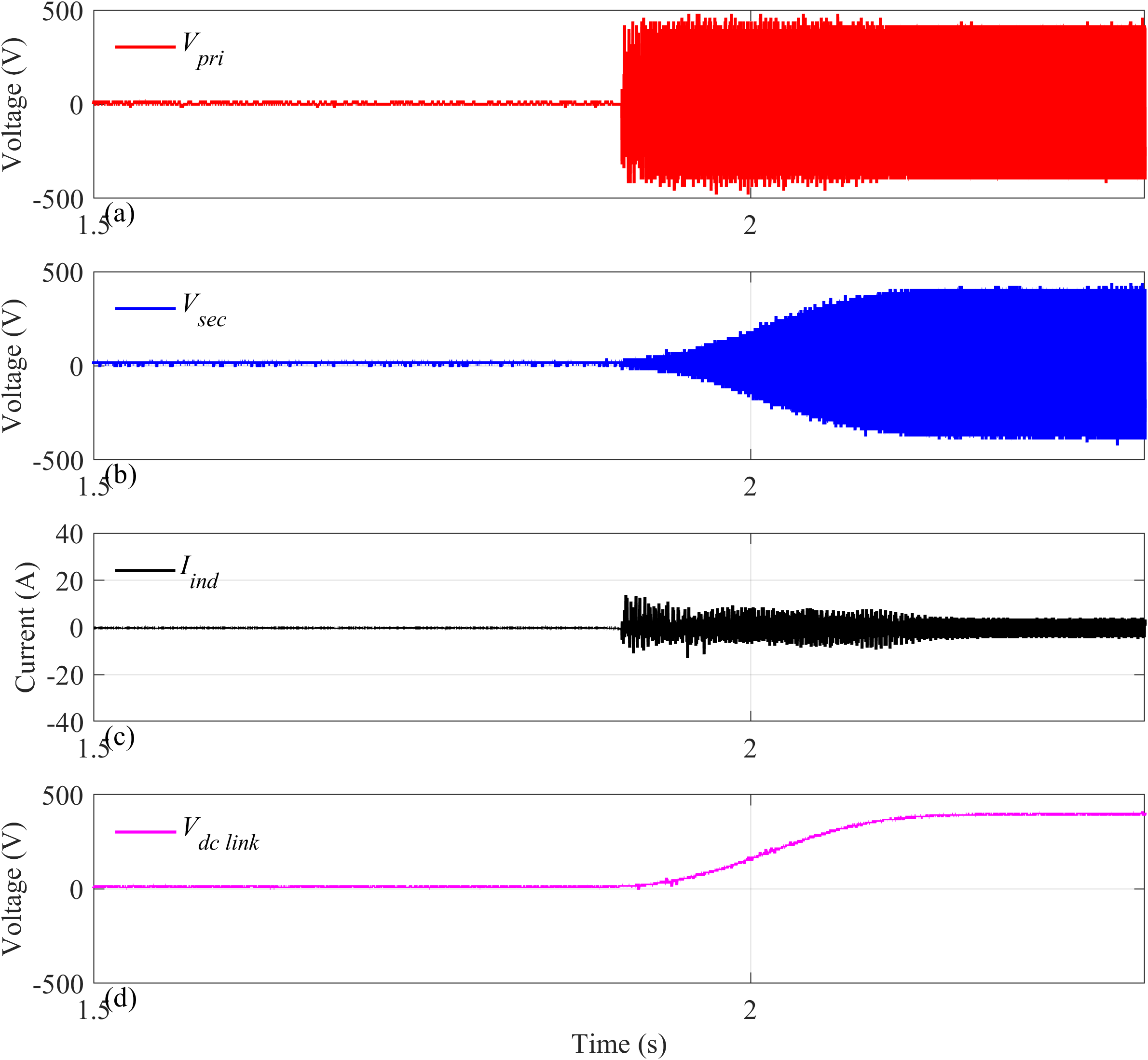}
	\caption{DAB behavior under proposed variable dead time-based startup process at 400V}\label{FIG_11}
\end{figure}

\begin{figure}[htbp]\centering
	\includegraphics[width=8.5cm]{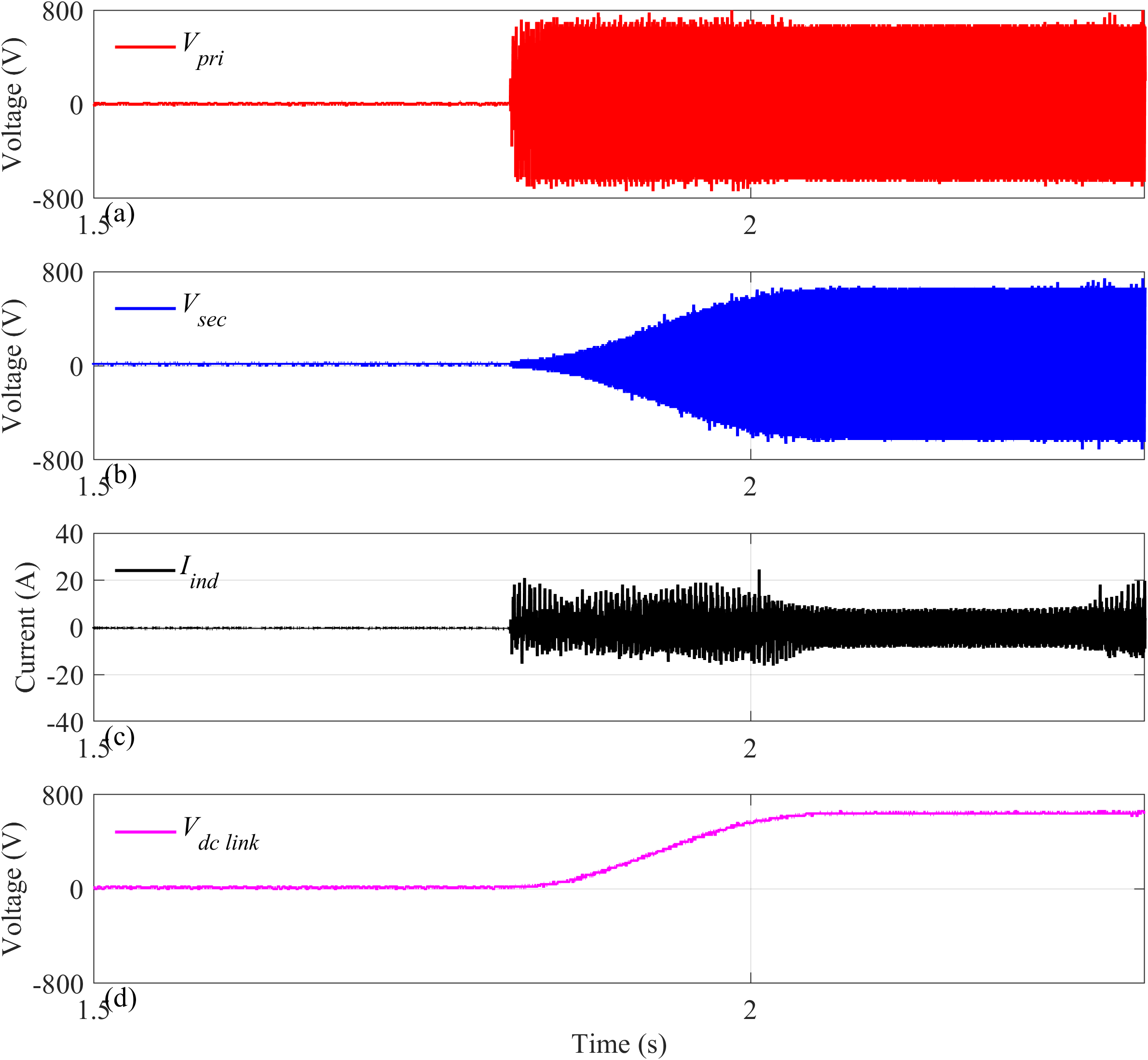}
	\caption{DAB behavior under proposed variable dead time-based startup process at 650V}\label{FIG_12}
\end{figure}

Figure 13 illustrates the effect of the dead-time variation rate on the output voltage response and the leakage inductor current. A higher rate of dead-time reduction results in a rapid voltage buildup, which can introduce sudden current surges and voltage transients. In contrast, a lower dead-time variation rate enables a smooth and gradual increase of the secondary-side output voltage, effectively mitigating transient behavior and ensuring stable soft-start operation.

\begin{figure}[htbp]\centering
	\includegraphics[width=8.5cm]{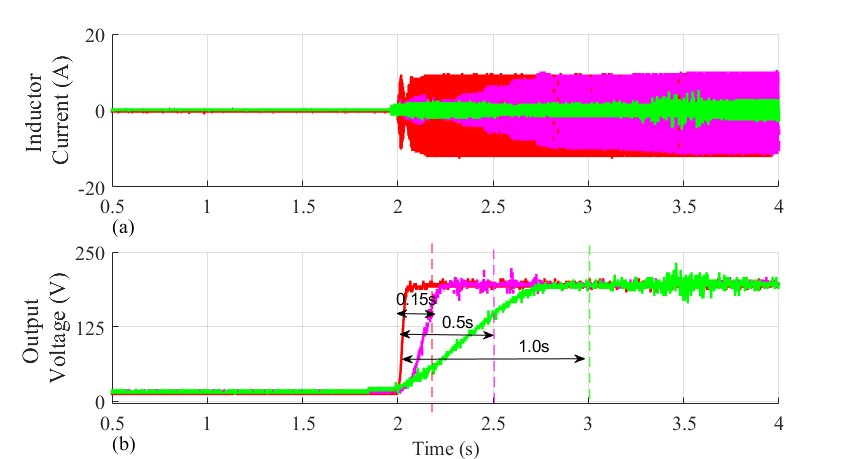}
	\caption{Proposed startup method at different startup rates.}\label{FIG_13}
\end{figure}

In contrast to conventional soft-start techniques, which often necessitate auxiliary pre-charge circuitry or complex fixed-duty ramping, the proposed method achieves superior voltage regulation and significantly attenuated inrush currents through software-defined modulation alone. This eliminates the need for additional hardware components, thereby increasing power density and reducing overall system cost. Furthermore, the experimental validation across diverse DC-link potentials confirms that the approach maintains high fidelity and operational robustness. Such characteristics make it an ideal candidate for a broad set of applications including integrated chargers, high-voltage EV traction inverters, and modular charging infrastructures where reliability and minimal transient stress is necessary.

\section{Conclusion}

This paper introduced a variable dead-time based soft-start methodology designed to mitigate the inherent transient limitations of conventional Dual Active Bridge (DAB) startup sequences. In traditional hard-start operations, the abrupt application of phase-shift or fixed dead-time leads to uncontrolled energy transfer, manifesting as severe secondary-side voltage overshoot and high-magnitude inrush currents that stress the transformer and power semiconductors. Unlike existing techniques, such as phase-shift modulation or single-cycle pulse-width control, the proposed method utilizes the dead-time as a primary control variable to achieve a monotonic and regulated startup.
By initiating the sequence with a dead-time approaching the switching period and linearly ramping it to the nominal value, the strategy effectively regulates the volt-second balance across the leakage inductance. This prevents current saturation and ensures a stable charging profile for the secondary-side capacitors. Both simulation and experimental validation on a 15 kW hardware prototype confirmed that this approach eliminates voltage overshoot and significantly attenuates inrush transients across various DC-link potentials. Ultimately, this variable dead-time strategy provides a robust, hardware-agnostic, and computationally efficient solution that enhances the reliability of DAB converters in high-power applications without increasing system complexity.

\section*{Acknowledgment}
This material is based upon work supported by the U.S. Department of Energy's Office of Energy Efficiency and Renewable Energy (EERE) under the Vehicle Technologies Office Award Number DE-EE0011171. The views expressed herein do not necessarily represent the views of the U.S. Department of Energy or the United States Government.


\bibliographystyle{Bibliography/IEEEtranTIE}


\vspace{1.2cm}
.

\end{document}